\documentclass{raa}            

\usepackage{graphicx,times}             
\input{epsf.sty}                        
\input{psfig.sty}                       

\begin{document}

   \title{A Model of Low-Frequency Quasi-Periodic Oscillations in Black Hole X-Ray Binaries
}

   \volnopage{Vol.0 (200x) No.0, 000--000}      
   \setcounter{page}{1}          

   \author{Zhi-Yun  Wang
      \inst{1,2}
   \and Chang-Yin  Huang
      \inst{2,3}
   \and Ding-Xiong  Wang
      \inst{2}$^{*}$
    \and Jiu-Zhou  Wang
      \inst{2}
   }

   \institute{School of Physics and Electronic Engineering, Xiangfan University, Xiangyang 441053, China; \\
        \and
             School of Physics, Huazhong University of Science and Technology, Wuhan 430074, China\\
        \and
             School of Mathematics and Statistics, Huazhong University of Science and Technology, Wuhan 430074, China\\
             {$^{*}$dxwang@mail.hust.edu.cn}\\
}

   \date{Received 2012 month day; accepted 2012 month day}

\abstract{ A model of low-frequency quasi-periodic oscillations
(LFQPOs) of black hole X-ray binaries (BHXBs) is proposed based on
the perturbed magnetohydrodynamic (MHD) equations of accretion disk. It
turns out that the LFQPOs frequencies of some BHXBs can be fitted
by the frequencies of the toroidal Alfv\'{e}n wave oscillation
corresponding to the maximal radiation flux. In addition, the
positive correlation of the LFQPO frequencies with the radiation
flux from accretion disk is well interpreted.
 }

   \authorrunning{Z. Y. Wang, C. Y. Huang, D. X. Wang \& J. Z. Wang}            
   \titlerunning{A Model of Low-Frequency Quasi-Periodic Oscillations in Black Hole X-Ray Binaries }  

  \maketitle

%
%
\section{Introduction}           
\label{sect:intro}

It is well known that the quasi-periodic oscillations (QPOs) have
been widely observed in some X-ray binaries, whose compact object
is either a neutron star or a black-hole, and QPOs play an
essential role as a potentially important tool for studying the
strong gravitational field and understanding the physical
processes of X-ray states (Done, Gierlinski \& Kubota~\cite{dg07}).
Among $\sim$20 black hole X-ray binaries (BHXBs), the
low-frequency QPOs (LFQPOs) with frequency ranging from a few $m$Hz
to 30Hz have been detected on one or more occasions for 14 systems in the hard state and the steep power-law
(SPL) state (McClintock \& Remillard~\cite{mr06}). Almost all
the LFQPOs have some common properties, remaining relatively
stable features for days or weeks usually.

A number of theoretical models have been proposed to explain the physical mechanisms of LFQPO.
Tagger and Pellat (\cite{tp99}) suggested
that LFQPOs observed in low-mass BHXBs can be interpreted by
accretion-ejection instability. Titarchuk \& Osherovich
(\cite{to00}) considered that the LFQPOs are caused by the global
disk oscillation in the direction normal to the disk, and these
oscillations arise from the gravitational interaction between the
central compact object and the disk. O'Neill et al. (\cite{or11})
proposed that LFQPOs are related to the quasi-periodic behavior in
global MHD dynamos. Cabanac et al. (\cite{ch10}) presented that an oscillating corona due to magneto-acoustic wave propagating in the corona, and produce multiple QPOs. Kato (\cite{k08}) proposed that the one-armed
c-mode low frequency oscillations of disk are one of possible
candidates of LFQPOs based on a resonantly-excited
disk-oscillation model. In addition, relativistic precession model was firstly presented by Stella and Vietri (\cite{sv98}), they suggested that LFQPOs are the result of some modulation of Lense-Thirring precession. Later, Schnittman et al. (\cite{sh06}) and Ingram \& Done et al. (\cite{id09},\cite{id12}) developed this model, and successfully explained the variability properties of BHXBs. However, there has been no consensus on the
physical nature of LFQPOs.

On the other hand, it has been detected that LFQPO frequencies exhibit a
strong positive correlation with disk flux from observations. For example, LFQPO frequencies of XTE J1550-564 (33 observations during 1998.9-1999.4), GRO J1655-40 (38 observations during 2005.2.17- 2005.3.6) and H1743-322 (20 observations during 2003.3.28-2003.5.22) display a roughly linear relation with disk
flux, when they are in the hard and intermediate states with
frequencies in the range 1-7Hz, 0.01-2.5Hz and 2-5Hz, respectively
(Remillard \& McClintock~\cite{rm06}; McClintock et
al.~\cite{mr09}; Sobczak et al.~\cite{sm00}; Shaposhnikov et
al.~\cite{ss07}). Similar quasi-linear relations of LFQPO
frequencies with disk flux have been found for GRS 1915+105 (32 observations during 1996.10.29-1998.10.7) in the
range 1-15Hz (Markwardt, Swank \& Taam~\cite{ms99}; Sobczak et
al.~\cite{sm00}; Muno, Remillard \& Morgan~\cite{mr01}) and for
XTE J1748-288 (92 observations during 1998.7.13-1998.9.26) in the range 20-30Hz (Revnivtsev, Trudolyubov \&
Borozdin ~\cite{rt00}). From the duration time of the above outbursts, we can find that this positive correlation between the LFQPO frequencies and disk flux is on long timescales. It is noted that the QPO observed in XTE J1550-564 during its 1998 outburst shows a correlation between absolute rms amplitude and mean source flux over timescales shorter than $\sim$\emph{3ks} (Heil,Vaughan \& Uttley ~\cite{hv11}), so this relation also holds on short timescales. However, the origin of the correlation
between LFQPO frequency and disk flux of BHXBs remains elusive.

If magnetic field is taken into account in accretion disk, a torsional Alfv\'{e}n wave can be generated by the rotational dragged of space (Koide et al.~\cite{ks02}). We consider a toroidal magnetic field existing in a rotational accretion disk, in which Alfv\'{e}n wave oscillation propagates along toroidal magnetic field lines due to the perturbation of radial velocity of the accreting matter. Thus the Alfv\'{e}n wave oscillation will influence the transformation of angular momentum and the radiation flux from the inner disk. Although some researchers suggested that the QPOs of low mass BHXBs arise from Alfv\'{e}n wave (Zhang~\cite{zh04}; Shi \& Li~\cite{sl10};Shi~\cite{sh11}), the association of Alfv\'{e}n wave
oscillation with LFQPOs has not been discussed.

In this paper, we adopt a binary system consisting of a Kerr black hole surrounded by weakly magnetized relativistic thin disk to derive the Alfv\'{e}n wave frequency based on the perturbed MHD equations, and the LFQPO frequency is fitted by the frequency of Alfv\'{e}n wave propagating in a specific circular orbit of the accretion disk, which corresponds to the maximum radiation flux. In addition, we propose a thin disk model with some corona floating above the disk to well interpret the positive correlation of the LFQPO frequencies with disk flux.

This paper is organized as follows. In section 2, we present a
description of our model, and derive the formula for Alfv\'{e}n
wave frequency and  radiation flux. In section 3, we fit the LFQPO
frequencies of several BHXBs, and fit the positive correlation of
LFQPO frequency with the disk flux for two BHXBs, XTE J1550-564
and GRO 1655-40. Finally, in section 4, we discuss the results
obtained in our model.


\section{MODEL DESCRIPTION}
\label{sect:Obs}
\subsection{Alfv\'{e}n Wave Oscillations in Accretion Disk}

We consider a geometrically thin, optically thick,
non-self-gravitating perfect fluid disk, which is magnetized and
isothermal, and rotating around a BH. It is assumed that the magnetic pressure is much less than the total pressure($p_{mag}\ll p$), which is a good approximation to a thin Keplerian disk (Li ~\cite{li02}). The disk dynamics is governed by the ideal MHD equations given as follows.
\begin{equation}
  \frac{\partial \rho}{\partial t}+\nabla\cdot(\rho\mathbf{v})=0,
\label{eq:1}
\end{equation}
\begin{equation}
  \frac{\partial\mathbf{v}}{\partial t}+\mathbf{v}\cdot\nabla\mathbf{v}+\frac{1}{\rho}\nabla p+\nabla \Phi+\frac{1}{4\pi\rho}\mathbf{B}\times(\nabla\times\mathbf{B})=0.
\label{eq:2}
\end{equation}

The magnetic field $\mathbf{B}$ satisfies the induction equation
in the MHD approximation, and it reads
\begin{equation}
  \frac{\partial\mathbf{B}}{\partial t}-\nabla\times(\mathbf{v}\times\mathbf{B})=0,
\label{eq:3}
\end{equation}
and
\begin{equation}
  \nabla\cdot\mathbf{B}=0.
\label{eq:4}
\end{equation}

The quantities $\rho$, $\mathbf{v}$ and $p$ in equations (1) and (2)
denote mass density, velocity and pressure of plasma,
respectively. The Pseudo-Kerr potential (Mukhopadhyay ~\cite{m02}) $\Phi$ is adopted to simulate the effects of general relativity, and it reads
\begin{equation}
  \nabla\Phi=-\frac{c^{4}(r^{2}-2a_{\ast}\sqrt{r}+a_{\ast}^{2})^{2}}{GMr^{3}[\sqrt{r}(r-2)+a_{\ast}]^{2}},
\label{eq:5}
\end{equation}
where $r$ is the distance between the plasma to center of BH ($r=R/R_{g}$), and $R_{g}\equiv GM/c^{2}$ is the gravitational radius. The quantities $M$,$G$ and $c$ denote respectively the BH mass, gravitational constant and speed of light, and $J$ and $a_{\ast}=J/(GM/c^{2})$ represent the BH angular momentum and the dimensionless spin, respectively.

The Keplerian angular velocity in the frame of Pseudo-Kerr potential is derived by Shafee et al. (\cite{sn08}) as follow,
\begin{equation}
  \Omega_{k}=\frac{c^{3}(r^{2}-2a_{\ast}\sqrt{r}+a_{\ast}^{2})}{GMr^{2}[\sqrt{r}(r-2)+a_{\ast}]}.
\label{eq:6}
\end{equation}

The perturbed physical quantities in MHD equations can be written as
\begin{equation}
  p=p_{0}+p',\rho=\rho_{0}+\rho',\mathbf{v}=\mathbf{v}_{0}+\mathbf{v}',\mathbf{B}=\mathbf{B}_{0}+\mathbf{B}',
\label{eq:7}
\end{equation}
where the subscript `0' and superscript `$'$' denote the equilibrium and perturbation values, respectively.
Assuming that
all the perturbations are small enough, i.e., $\rho_{0}\gg\rho'$
, $p_{0}\gg p'$, $v_0\gg v'$, $B_0\gg B'$, we can neglect the
products of them in second and higher orders. Substituting
equations (4) and (7) into equations (1)-(3), we have the
linearized MHD perturbation equations as follows,
\begin{equation}
  \frac{\partial \rho'}{\partial t}+\rho_{0}(\nabla\mathbf{v'})+(\nabla\rho')\cdot\mathbf{v}_{0})=0,
\label{eq:8}
\end{equation}
\begin{equation}
  \frac{\partial\mathbf{v'}}{\partial t}+\mathbf{v_{0}}\cdot(\nabla\mathbf{v'})+\frac{c_{s}^{2}}{\rho_{0}}\nabla\rho'+\frac{\mathbf{B_{0}}}{4\pi\rho_{0}}
\times(\nabla\times\mathbf{B'})=0,
\label{eq:9}
\end{equation}
\begin{equation}
  \frac{\partial\mathbf{B'}}{\partial t}-\nabla\times(\mathbf{v'}\times\mathbf{B_{0}})=0,
\label{eq:10}
\end{equation}
From the vertical equilibrium assumption of accretion disk, the half-height(\emph{H}) of the disk can be written as (Mukhopadhyay ~\cite{m03}),
\begin{equation}
  H=c_{s}R^{1/2}F^{-1/2}=c_{s}/\Omega_{k},
\label{eq:11}
\end{equation}
where $c_{s}=\sqrt{p/\rho}$ is the sound speed.

Incorporating equations (8)-(10), we have the equation for the
velocity perturbation as follows,
\begin{equation}
 \frac{\partial^{2}\mathbf{v'}}{\partial t^{2}}+\mathbf{v_{0}}\cdot\nabla(\frac{\partial\mathbf{v'}}{\partial t})+c_{s}^{2}\nabla(\nabla\cdot\mathbf{v'})+\mathbf{v_{A}}\cdot{\nabla\times[\nabla\times(\mathbf{v'}\times\mathbf{v_{A}})]}=0,
\label{eq:12}
\end{equation}
where $\mathbf{v_{A}}=\frac{\mathbf{B_{0}}}{\sqrt{4\pi\rho_{0}}}$
is Alfv\'{e}n velocity, being defined as
\begin{equation}
 v_{A}=\frac{|\mathbf{B_{0}}|}{\sqrt{4\pi\rho_{0}}}=\sqrt{\beta}c_{s}=\sqrt{\beta}H\Omega_{k},
\label{eq:13}
\end{equation}
where $\beta=p_{mag}/p$ is defined as the ratio of magnetic
pressure to total pressure.

In order to fit LFQPO frequencies by invoking Alfv\'{e}n wave
propagating in an accretion disk, we express the perturbed
velocity as follows,
\begin{equation}
 \mathbf{v'}=\mathbf{v'}e^{i(\mathbf{k}\cdot\mathbf{\xi}-\omega t)},
\label{eq:14}
\end{equation}
where $\omega$, $\mathbf{\xi}$ and $\mathbf{k}$ are the perturbation frequency, the displacement vector and the wavenumber vector, respectively. In cylindrical coordinates the
wavenumber vector is
$\mathbf{k}=k_{r}\mathbf{\hat{r}}+k_{\varphi}\mathbf{\hat{\varphi}}+k_{z}\mathbf{\hat{z}}$,
where $k_{r},k_{\varphi},k_{z}$ represent the radial, azimuthal
and vertical components, respectively. Substituting equation (14)
into equation (12), we have
\begin{equation}
 -\omega^{2}\mathbf{v'}-\omega k\mathbf{v_{0}}\cdot\mathbf{v'}+(c_{s}^{2}+v_{A}^{2})(\mathbf{k}\cdot\mathbf{v'})\mathbf{k}+(\mathbf{v_{A}}\cdot\mathbf{k})\cdot
 [(\mathbf{v_{A}}\cdot\mathbf{k})\mathbf{v'}-(\mathbf{v_{A}}\cdot\mathbf{v'})\mathbf{k}-(\mathbf{k}\cdot
 \mathbf{v'})\mathbf{v_{A}}]=0.
\label{eq:15}
\end{equation}

Considering that Alfv\'{e}n wave is a transverse wave,
transporting along the magnetic field line, we have
$\mathbf{k}\|\mathbf{B_{0}}\|\mathbf{v_{0}}\perp\mathbf{v'}$, and
equation (14) is simplified as
\begin{equation}
 k^{2}v_{A}^{2}-\omega^{2}=0,
\label{eq:16}
\end{equation}
where the symbols $\|$ and $\perp$ denote parallel and vertical
directions, respectively. Thus the angular velocity of the
perturbation is related to the Alfv\'{e}n velocity $v_{A}$ as
follows,
\begin{equation}
 \omega=kv_{A}=k\sqrt{\beta}c_{s}.
\label{eq:17}
\end{equation}

For a thin magnetized accretion disk, Armitage and Natarajan (\cite{an99}) presented that the dominant field component is toroidal with saturation occurring when $p_{mag}\ll p$. So we assume that the unperturbed flow is axisymmetric, and only toroidal magnetic field exists in equilibrium state in the accretion disk,i.e., $\mathbf{B_{0}}=B_{\varphi}\hat{\varphi}$.  Thus a small perturbation of radial velocity of fluid due to accretion of black hole gives rise to the oscillation of the magnetic field $B_{\varphi}$, and results in
the Alfv\'{e}n wave oscillation transporting in the toroidal
direction in the disk. For the characteristic wavelength
$\lambda\sim R$, the toroidal wave number
  $k_{\varphi}\sim2\pi/R$ (Shi,~\cite{sl10}), and the angular velocity of the perturbation can be written as
\begin{equation}
 \omega=k_{\varphi}\sqrt{\beta}c_{s}=2\pi\frac{H}{R}\sqrt{\beta}\Omega_{k}.
\label{eq:18}
\end{equation}

Inspecting equation (18) we find that $\omega$ is much less than
the Keplerian angular velocity $\Omega_{k}$ in the thin disk due
to the disk scale height $H$ being much less than disk radius $R$,
and this perturbation frequency provides a possibility for fitting LFQPO frequencies of BHXBs.

\subsection{Relation between LFQPO Frequency and Perturbation Frequency}

First of all, we intend to clarify the relation between LFQPO
frequency and the perturbation frequency. The disk angular
velocity can be regarded as the Keplerian angular velocity, i.e.,
$\Omega=\Omega_{k}$, provided that the radial magnetic force can
be neglected. This result can be realized if only toroidal
magnetic field exists without vertical electric current in the
thin disk.

The rate of energy generation per unit area of one side of the
disk is given by (Shakura \& Sunyav ~\cite{ss73}; Novikov \& Thorne ~\cite{nt73}; Gierli\'{n}ski et al.~\cite{gz99};  Shafee et al.~\cite{sn08}),
\begin{equation}
 Q(R)=-RH\alpha p\frac{d\Omega_{R}}{dR}=-\frac{\dot{M}(R^{2}\Omega_{R}-R_{ms}^{2}\Omega_{ms})}{4\pi R}\cdot\frac{d\Omega_{R}}{dR},
\label{eq:19}
\end{equation}
where $\dot{M}$ and $\alpha$ are respectively the mass accretion rate and viscosity parameter, $\Omega_{R}$ and $\Omega_{ms}$ denote respectively the angular velocity at $R$ and inner edge of disk. The radius $R_{ms}$ of the  innermost stable circular orbit (ISCO) of accretion disk varies with $a_{\ast}$ in Pseudo-Kerr potential, Mukhopadhyay (\cite{m02}) presented that $r_{ms}$ ($r_{ms}=R_{ms}/R_{g}$) satisfies the following equation
\begin{equation}
 -3a_{\ast}^{4}+14a_{\ast}^{3}\sqrt{r_{ms}}+(r_{ms}-6)r_{ms}^{3}+6a_{\ast}r_{ms}^{3/2}(r_{ms}+2)-2a_{\ast}^{2}r_{ms}(r_{ms}+11)=0.
\label{eq:20}
\end{equation}

It is easy to find that ISCO moves to the BH with the increasing $a_{\ast}$ as shown in Fig.1.Thus the local radiation flux $F(r)$ can be written as:
\begin{equation}
F(r)=\frac{\dot{M}c^{6}}{8\pi G^{2}M^{2}r^{4}}g(r,r_{ms},a_{\ast}),
\label{eq:21}
\end{equation}
where we have $r=R/R_{g}$, and
\begin{eqnarray}
g(r,r_{ms},a_{\ast}) & = &
(2r^{2}\sqrt{r_{ms}}-r^{2}\sqrt{r_{ms}^{3}}-r^{2}a_{\ast}+2a_{\ast}\sqrt{rr_{ms}^{3}}
-a_{\ast}^{2}\sqrt{r_{ms}^{3}}+r_{ms}^{2}\sqrt{r^{3}}-2r_{ms}^{2}\sqrt{r} \nonumber \\
&  & +r_{ms}^{2}a_{\ast}-2a_{\ast}\sqrt{r^{3}r_{ms}}+a_{\ast}^{2}\sqrt{r^{3}})(12\sqrt{r^{5}}a_{\ast}
-3r^{4}+16a_{\ast}^{2}r-16a_{\ast}\sqrt{r^{3}}-7r^{2}a_{\ast}^{2} \nonumber \\
&  & +2r^{3}-4a_{\ast}^{3}\sqrt{r})/[\sqrt{r}(\sqrt{r^{3}}-2\sqrt{r}+a_{\ast})^{3}(\sqrt{r_{ms}^{3}}
-2\sqrt{r_{ms}}+a_{\ast})].
\label{eq:22}
\end{eqnarray}

By using equation (20) and (21), we plot the curves of local flux $F(r)$ versus $r$ as shown in Fig.1, and we find that the radiation flux is dominantly produced in the inner disk, and it varies non-monotonically with $r$, attaining its peak value at $r_{d}$ close to ISCO. In addition, as shown in Fig.1, we find that the peak value of $F(r)$ is greater, and the location of $r_{d}$ is closer to ISCO for a greater BH spin.
\begin{figure}
\centering
    \includegraphics[width=80mm,height=60mm]{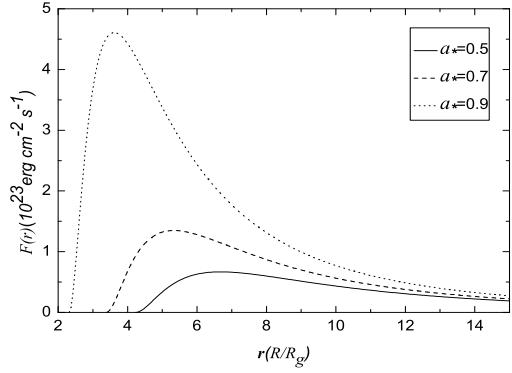}
  \caption{The curves of local radiation flux $F(r)$ versus r with the given values of $a_{\ast}$. \
      ($M=10M_{\odot}$,$\dot{M}=0.1\dot{M}_{Edd}$,$\dot{M}_{Edd}$ is the Eddington accretion rate.)}
  \label{Fig:fig1}
\end{figure}

Inspecting equations (18)-(21), we can find that the radiation flux is directly caused by the angular momentum transport. For magnetized thin accretion disk, numerical simulations have shown that the angular momentum transport is dominated by Maxwell stresses (Hawley et al.~\cite{hg95}; Brandenburg et al.~\cite{bn95}; Stone et al.~\cite{sh96}), and the magnetic contribution to viscosity parameter $\alpha$ exceeds fluid stresses by an order of magnitude, so that $\alpha\simeq\alpha_{magnetic}$ (Armitage and Natarajan~\cite{an99}). So in low mass X-ray binaries the high energy X-ray radiation is generated mainly from the interaction of the plasma with the magnetic field, and the change of the magnetic field could modulate the X-ray flux (Shi~\cite{sh11}). Thereby the oscillations of the toroidal Alfv\'{e}n wave may lead to the QPOs. Since the strongest influence of the perturbation on luminosity corresponds to the maximum radiation flux $F_{max}(r=rd)$, the perturbation frequency at $r_{d}$ can be regarded as the LFQPO frequency.

\section{FITTING LFQPOs OF BHXBs}
\label{sect:fitting}

In this section we intend to fit the LFQPOs of several BHXBs based on the above relation between LFQPO frequency and the perturbation frequency. In addition, the positive correlation between LFQPO frequency and disk flux can be well interpreted based on our model.

We consider a geometrically thin, optically thick accretion disk with corona floating above given by Gierli\'{n}ski et al. (\cite{gz99}). In the inner region of weak magnetic fields, the gas pressure and magnetic pressure can be neglected, and the total pressure is
\begin{equation}
 p\simeq p_{rad}=\frac{4\sigma}{3c}T_{c},
\label{eq:23}
\end{equation}
where $\sigma$ is Stefan-Boltzmann constant, and the
quantity $T_{c}$ is the temperature of the central disk, which is
determined by the equilibrium between the radiation cooling in the
vertical direction and the energy generated by the viscous
dissipation. We assume that a fraction $q$ of the total energy generated by the viscous process is emitted from the disk, and the remainder is dissipated in the corona. This energy equation is written as (Frank, King \&
Raine~\cite{fk02}; Gierli\'{n}ski et al. \cite{gz99})
\begin{equation}
 \frac{4\sigma}{3\tau}T_{c}^{4}=qF(r).
\label{eq:24}
\end{equation}

The parameter $\tau$ in equation (24) is the opacity for electron
scattering, and it reads
\begin{equation}
 \tau\cong\rho H\sigma_{T}/m_{p},
\label{eq:25}
\end{equation}
where $\sigma_{T}$ and $m_{p}$ denote the Thomson cross-section
and proton mass, respectively. Incorporating equation (11) with
equations (23)-(25), we obtain
\begin{equation}
 H=\frac{\sigma_{qT}}{m_{p}c(\Omega_{k})^{2}}F(r).
\label{eq:26}
\end{equation}
The Eddington accretion rate $\dot{M}_{Edd}$ can be written as
\begin{equation}
 \dot{M}_{Edd}=2\pi R_{ms}m_{p}c/\eta\sigma_{T},
\label{eq:27}
\end{equation}
where $\eta$ is the accretion efficiency, whose value is 0.1 for black hole. Substituting equations (6), (21) and (27) into equation (26), we have
\begin{equation}
 H=3.7\times10^{5}m_{BH}\dot{m}q\frac{r_{ms}(\sqrt{r^{3}}-2\sqrt{r}+a_{\ast})^{2}}{(2a_{\ast}\sqrt{r}-r^{2}
 -a_{\ast}^{2})^{2}}g(r,r_{ms},a_{\ast}),
\label{eq:28}
\end{equation}
where we define $m_{BH}\equiv M/M_{\odot}$ and
$\dot{m}\equiv \dot{M}/\dot{M}_{Edd}$.

Combining equations (6), (18) and (28), we have the perturbation frequency as a function of disk radius $r$, and we derive the expression for LFQPO frequency by setting $r=r_{d}$ as follows,
\begin{equation}
 \nu_{QPO}=\frac{\omega}{2\pi}=5.1\times10^{5}\frac{\sqrt{\beta}\dot{m}q}{m_{BH}}\frac{r_{ms}(\sqrt{r_{d}^{3}}-2\sqrt{r_{d}}
 +a_{\ast})}{r_{d}^{3}(2a_{\ast}\sqrt{r_{d}} -r_{d}^{2}-a_{\ast}^{2})}g(r_{d},r_{ms},a_{\ast}).
\label{eq:29}
 \end{equation}

From equation (29) we find that the LFQPO frequency is proportional to accretion rate $\dot{m}$, but it is inversely proportional to the BH mass $m$. Based on the above results we can fit some LFQPO frequencies of
several BHXBs with the given BH mass by adjusting the mass
accretion rate and the intensity of magnetic field $\beta$, and the main results are summarized as follows.

First, the theoretical values of the LFQPO frequencies are in
accordance with the observed ones with appropriate values of
$\dot{m}$ and $\beta$. We assume that the dissipation in corona is neglected ($q=1$), and $\beta=p_{mag}/p=0.01$, and the fitting results are listed in Table 1.

\begin{table}
\begin{center}

      \caption[]{Fitting LFQPO Frequencies of X-ray BHBs.}
\label{Tab:publ-works}
 \begin{tabular}{cccccc}
  \hline\noalign{\smallskip}
     source        &  $\nu_{QPO}^{a}$  &  $m_{BH}^{b}$  & $a_{\ast}^{b}$   & $\beta=p_{mag}/p$  & $\dot{m}$ \\
  \hline\noalign{\smallskip}
   GRO J1665$-$40  & 0.1$-$28          &  $6.2$         &   0.7            &                    &  0.00086$-$0.24 \\
   XTE J1550$-$564 & 0.1$-$10          &  $9.1$         &   0.34           &   0.01             &  0.0023$-$0.23  \\
   GRS 1915$+$105  & 0.001$-$10        &  $15.0$        &   0.975          &                    &  0.00007$-$0.07 \\
   4U 1543$-$47    &  7                &  $9.4$         &    0.8           &                    &    0.073   \\
  \noalign{\smallskip}\hline
\end{tabular}

Notes: $^{a}$Remillard \& McClintock (\cite{rm06});  $^{b}$Narayan \&
McClintock (\cite{nm11})
\end{center}
\end{table}

Second, the observed strong positive correlation between the LFQPO frequencies and the disk flux can be interpreted very well based on our model.

We assume that the disk locally emits a blackbody spectrum, of which a fraction $1-p_{sc}$ is scattered in the corona, and the disk luminosity is (Gierli\'{n}ski et al.~\cite{gz99})
\begin{equation}
 L_{s}=p_{sc}q\eta\dot{M}c^{2}=\frac{2\pi D^{2}F_{s}}{\cos i},
\label{eq:30}
\end{equation}
where $i$ and $D$ are the inclination of the disk and the source
distance to the observer, respectively. The quantity$(1-p_{sc})$
is the fraction of the disk emission which is not scattered by the
corona, and $F_{s}$ is the observed disk flux. Combining equations (29)
and (30), we obtain the relation between the LFQPO frequency and
the observed disk flux $F_{s}$ as follows.
\begin{equation}
\nu_{QPO}=2.5\times10^{-33}\frac{\sqrt{\beta}F_{s}D^{2}}{m_{BH}^{2}p_{sc}\cos i}\frac{r_{ms}(\sqrt{r_{d}^{3}}-2\sqrt{r_{d}}
 +a_{\ast})}{r_{d}^{3}(2a_{\ast}\sqrt{r_{d}} -r_{d}^{2}-a_{\ast}^{2})}g(r_{d},r_{ms},a_{\ast}).
\label{eq:31}
\end{equation}

From equation (31) we find that LFQPO frequency has a positive correlation with the observed disk
 flux. For XTE J1550-564 (outburst in 1998) and GRO J1655-40 (outburst in early 2005). The values
 of $p_{sc}$ are about 15\% (Sobczak et al.~\cite{sm00}) and 10\% (Shaposhnikov et al.~\cite{ss07}), respectively, and they keep constant in hard state. Therefore their disk fluxes have a linear relation with LFQPO frequencies.

 For a  given source, the values of $r_{ms}$ and $r_{d}$ can be obtained by resolving equations (20) and (21), and the values of the parameters, such as $m_{BH}, a_{\ast}, D, i$, and $p_{sc}$)are available in the literatures. Substituting all the parameters into the equation (31), we can derive the LFQPO frequency corresponding to disk flux $F_{s}$ and the magnetic field parameter $\beta$ as follows,
 \begin{equation}
 \nu_{QPO}=f(\beta)F_{s},
 \label{eq:32}
 \end{equation}
 where $f(\beta)$ is the function of $\beta$. The measurement values of LFQPO frequency and disk flux are $\nu_{i}$  and $F_{i}$ ($i=1,2,\cdots,N$, the number of the observation data), of which the errors are $\epsilon_{\nu i}$ and $\epsilon_{Fi}$.We adopt the Nukers' estimate (Tremaine et al.~\cite{tg02}) based on the following minimizing
 \begin{equation}
 \chi^{2}\equiv \sum\limits_{i=1}^{N}\frac{(\nu_{i}-f(\beta)F_{i})^{2}}{\epsilon_{\nu i}^{2}+f^{2}(\beta)\epsilon _{Fi}^{2}}.
 \label{eq:33}
 \end{equation}

 Using the Nukers' estimate, we can derive the best fitting parameter $\beta$ for XTE J1550-564 and GRO J1655-40, which are presented in Table 2, and the values of reduced $\chi^{2}$ per degree of freedom are less than 2 for both sources, indicating that LFQPO frequency has approximate linear relation with disk flux as shown in Figs.2 and 3 for XTE J1550-564 and GRO J1655-40, respectively.

 \begin{table}
\begin{center}

      \caption[]{LFQPO Frequencies Fitting Parameters of XTE J1550-564 and GRO J1655-40.}
\label{Tab:publ-works}
 \begin{tabular}{cccccccccc}
  \hline\noalign{\smallskip}
     source      &Input   &          &        &      &        &Output  &       &Fitting parameter&$\chi^{2}(dof)$\\
                 &$m_{BH}$&$a_{\ast}$&$D(kpc)$&$i$   &$p_{sc}$&$r_{ms}$&$r_{d}$&$\beta=p_{mag}/p$&              \\
  \hline\noalign{\smallskip}
  XTE J1550$-$564&9.1     & 0.34     & 4.38   &$75^{0}$& 0.15   & 4.83   & 7.64  &$0.0105\pm0.0003$& 44.9(32)      \\
  GRO J1665$-$40 &6.3     & 0.7      & 3.2    &$70^{0}$&  0.1   & 3.39   & 5.33  &$0.0159\pm0.0005$& 61.2(37)      \\
  \noalign{\smallskip}\hline
\end{tabular}

Notes:The values of input parameters ($m_{BH},a_{\ast},D,i$) are adopted from Narayan \
\& McClintock (\cite{nm11}).\\
The quantities of $r_{ms}$ and $r_{d}$ are computed by using equations (20) and (21). \  \  \  \  \  \  \  \  \  \  \  \  \  \  \  \  \  \  \  \  \  \  \  \
\end{center}
\end{table}

\begin{figure}
\centering
    \includegraphics[width=80mm,height=60mm]{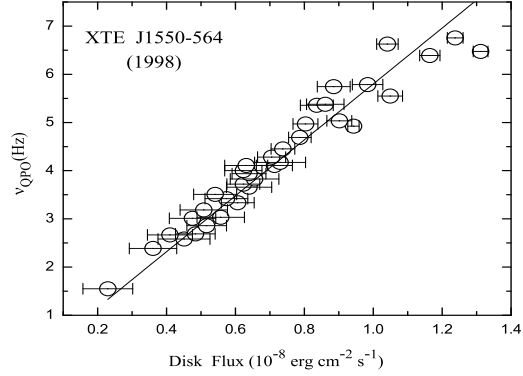}
   \caption{Fitting the relation between the disk flux and LFQPO frequency for XTE J1550-564. The observation data are taken from Fig. 2 Of SobcZak et al. (2000), while the flux errors are corrected based on McClintock et al. (2009).}
  \label{Fig:fig2}
\end{figure}
\begin{figure}
\centering
    \includegraphics[width=80mm,height=60mm]{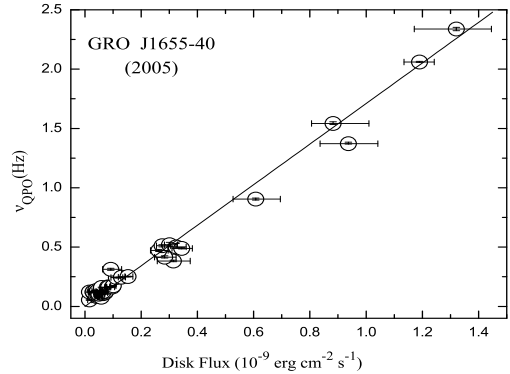}
   \caption{Fitting the relation between the disk flux and LFQPO frequency for GRO J1655-40. ($m_{BH}=6.2$, $\beta=0.6$, $D=3.2 \rm kpc$, $i=70^{0}$. The observation data are taken from Fig. 15 of Shaposhnikov et al. (2007).}
  \label{Fig:fig3}
\end{figure}

\section{DISCUSSION}
\label{sect:discussion}

In this paper, we propose that the LFQPOs in BHXBs can be
interpreted by invoking the toroidal Alfv\'{e}n wave oscillations
located at the disk radius with the maximal radiation flux. It
turns out that the LFQPO frequencies of several BHXBs can be well
fitted based on our model. In addition, the positive correlation
between the LFQPO frequencies and the disk flux of XTE J1550-564
and GRO J1655-40 are also well fitted. According to the argument
given in sections 2 and 3, we find that the positive correlation between LFQPO frequency and the disk flux can be understood, because both the perturbation frequency of the toroidal Alfv\'{e}n wave and LFQPO frequency increase with the increasing accretion rate. It is noted that this correlation only holds in hard stats, when the disk fraction keeps constant. Thus our model provides an explanation for the fact that LFQPOs are primarily observed in the power law part of spectra of BHXBs (Zycki \& Sobolewska~\cite{zs05}).

In this paper, we adopt a thin disk with some corona floating above, which is different from the corona interior to a truncated disk as given by Done, Gierlinski \& Kubota (\cite{dg07}). The truncated disk is not compatible with our model because it is difficult to determine the location of maximal radiation flux. If Alfv\'{e}n wave of truncated disk is used to interpret QPOs, a certain appropriate location of Alfv\'{e}n wave propagating must be found out, e.g. Shi (\cite{sh11}) used Alfv\'{e}n wave oscillation at transition radius to explain HFQPOs of low mass X-ray binaries.

Although LFQPOs are interpreted successfully by invoking perturbed Alfv\'{e}n wave oscillation, it seems difficult to explain high Frequency QPOs (HFQPOs) of BHXBs. This could imply that the physical origins of LFQPOs and HFQPOs are different. HFQPOs of BHXBs have been successfully explained by the magnetic hot spots model (Wang, et al.~\cite{wang03},~\cite{wang05}) and resonance mode (Abramowicz and Kluzniak~\cite{ak01}; Huang, et al.~\cite{hgw10}), and the HFQPO frequencies are sensitive to BH mass and spin, while they seems not directly related to the accretion rate and disk flux. It is noted that HFQPOs may be triggered by the instability of accretion disk oscillation (Tagger et al.~\cite{tv06},~\cite{ct01}). We also notice that the instability occurs in the toroidal Alfv\'{e}n wave oscillation, and the oscillation frequency is comparable to the HFQPO frequency, provided that the accretion rate and the magnetic field are great enough. This result motivates us to explore the relation of HFQPOs and the instability of Alfv\'{e}n wave oscillation in our future work.

\begin{acknowledgements}
This work is supported by the NSFC (grants 11173011, 11143001, 11103003 and 11045004), the National Basic Research Program of China (2009CB824800) and the Fundamental Research Funds for the Central Universities (HUST: 2011TS159).
\end{acknowledgements}

\label{lastpage}

\end{document}